\documentclass[conference]{IEEEtran}
\IEEEoverridecommandlockouts

\usepackage{cite}
\usepackage{amsmath,amssymb,amsfonts}
\usepackage{bm}
\usepackage{algorithm}
\usepackage{algpseudocode}
\usepackage{graphicx}
\usepackage{booktabs}
\usepackage{textcomp}
\usepackage{xcolor}
\usepackage{url}
\def\BibTeX{{\rm B\kern-.05em{\sc i\kern-.025em b}\kern-.08em
    T\kern-.1667em\lower.7ex\hbox{E}\kern-.125emX}}

\begin{document}

\title{A Polynomial-Decay and Pinhole-Imaging Whale Optimization Algorithm for UAV Relay Communication Deployment%
\thanks{$^{*}$Corresponding authors: Junhao Wei (p2312195@mpu.edu.mo) and Yapeng Wang (yapengwang@mpu.edu.mo).}}

\author{\IEEEauthorblockN{Zhenhong Peng}
\IEEEauthorblockA{\textit{College of Animal Science and Technology} \\
\textit{Zhongkai University of Agriculture and Engineering}\\
Guangzhou, China}
\and
\IEEEauthorblockN{Junhao Wei$^{*}$}
\IEEEauthorblockA{\textit{Faculty of Applied Sciences} \\
\textit{Macao Polytechnic University}\\
Macao SAR, China}
\and
\IEEEauthorblockN{Baili Lu}
\IEEEauthorblockA{\textit{Faculty of Applied Sciences} \\
\textit{Macao Polytechnic University}\\
Macao SAR, China}
\and
\IEEEauthorblockN{Yanxiao Li}
\IEEEauthorblockA{\textit{Faculty of Applied Sciences} \\
\textit{Macao Polytechnic University}\\
Macao SAR, China}
\and
\IEEEauthorblockN{Yifu Zhao}
\IEEEauthorblockA{\textit{Faculty of Applied Sciences} \\
\textit{Macao Polytechnic University}\\
Macao SAR, China}
\and
\IEEEauthorblockN{Haochen Li}
\IEEEauthorblockA{\textit{Faculty of Applied Sciences} \\
\textit{Macao Polytechnic University}\\
Macao SAR, China}
\and
\IEEEauthorblockN{Dexing Yao}
\IEEEauthorblockA{\textit{Faculty of Applied Sciences} \\
\textit{Macao Polytechnic University}\\
Macao SAR, China}
\and
\IEEEauthorblockN{Xu Yang}
\IEEEauthorblockA{\textit{Faculty of Applied Sciences} \\
\textit{Macao Polytechnic University}\\
Macao SAR, China}
\and
\IEEEauthorblockN{Yapeng Wang$^{*}$}
\IEEEauthorblockA{\textit{Faculty of Applied Sciences} \\
\textit{Macao Polytechnic University}\\
Macao SAR, China}}

\maketitle

\begin{abstract}
Unmanned aerial vehicle (UAV) relays deliver flexible, on-demand wireless coverage, but jointly tuning the position, altitude, transmit power and bandwidth of the relay is a non-convex, heavily constrained optimization task that easily traps swarm-based optimizers in poor local optima. We propose PWOA, a Polynomial-decay and Pinhole-imaging Whale Optimization Algorithm with three complementary improvements: (i) a Good Nodes Set (GNS) initialization that spreads the initial population uniformly across the search space; (ii) a polynomial nonlinear schedule for the convergence factor that prolongs early exploration and sharpens late exploitation; and (iii) a stagnation-triggered pinhole-imaging opposition-based learning (POBL) operator paired with an elite Gaussian local search, which together escape local optima while refining the leader. On a five-dimensional UAV relay deployment problem with five inequality constraints ($N{=}30$, $T{=}500$, 30 independent runs), PWOA simultaneously attains the lowest Best, Worst, Mean and standard deviation among PWOA, WOA, SCA and IPSO, cutting the mean by $1.4$--$18.5\%$ and the standard deviation by $15$--$87\%$ over the three baselines, and exhibits the fastest average convergence.
\end{abstract}


\section{Introduction}
\label{sec:intro}

UAV-mounted aerial relays provide rapid infrastructure-less coverage for disaster recovery, hot-spot offloading and tactical communication. Their deployment jointly determines the 3D position, transmit power and bandwidth under line-of-sight, capacity and energy constraints, yielding a non-convex black-box problem suited to meta-heuristics \cite{wei2024apupso,wei2026saga}.

The Whale Optimization Algorithm (WOA) \cite{mirjalili2016woa} is attractive because of its simple structure and few control parameters, but random initialization and the linear decay of the convergence factor $a$ can cause uneven coverage and premature exploitation. Variants such as RWOA \cite{wei2025rwoa}, LSEWOA \cite{wei2025lsewoa}, TSWOA \cite{wei2025tswoa}, LSWOA \cite{wei2025lswoa}, geometric WOA \cite{wei2026geometric}, CICDWOA \cite{wei2026cicdwoa}, and related hybrids or applications \cite{lu2025mrbmo,wei2026nawoa,li2026asksssa} show that initialization, perturbation and adaptive scheduling are decisive design choices.

PWOA is therefore designed as a role-coupled optimizer rather than a loose accumulation of WOA variants: GNS improves initial coverage, polynomial decay controls the exploration-to-exploitation transition, and stagnation-triggered POBL with elite search recovers the leader after stagnation. This operator--failure-mode mapping is consistent with UAV optimization studies such as landscape-aware bandit hyper-heuristics \cite{wei2026landscape} and GeoSSA \cite{wei2026geossa}, which show that operator design should match the search landscape.

Learning-based and model-driven paradigms are complementary. Dense-reward deep reinforcement learning has been used for UAV path planning \cite{zhou2025uav} and is attractive with repeated online decisions and sufficient training scenarios, while model-driven methods can offer structure under tractable assumptions. PWOA instead targets one-shot or limited-data relay deployment where the objective and constraints are directly evaluable but hard to convexify or differentiate.

The contributions are threefold: (i) a five-dimensional constrained UAV relay deployment formulation with five inequality constraints and a quadratic penalty function; (ii) PWOA, which combines GNS initialization, polynomial nonlinear decay and stagnation-triggered POBL with elite Gaussian local search; and (iii) a 30-run comparison against WOA, SCA and IPSO, where PWOA achieves the best Mean, Std and convergence speed.

\section{Problem Formulation}
\label{sec:problem}

We consider a UAV that hovers above a service area to relay communication between two ground nodes located at $\mathbf{p}_1=(1.5,2.0)$ km and $\mathbf{p}_2=(8.0,7.0)$ km. The decision vector
\begin{equation}
\mathbf{x} = [x,\ y,\ h,\ p_{tx},\ B]^\top
\end{equation}
encodes the horizontal position $(x,y)$ in km, altitude $h$ in metres, transmit power $p_{tx}$ in watts and bandwidth $B$ in MHz, with bounds
\begin{equation}
\mathbf{lb}=[0,0,80,1,1],\quad \mathbf{ub}=[10,10,400,8,20].
\end{equation}

\paragraph{Link model} The horizontal and full distances to the two ground nodes are
\begin{align}
d_{h,i} &= \|\,(x,y)-\mathbf{p}_i\,\|_2,\quad i\in\{1,2\}, \\
d_i &= \sqrt{d_{h,i}^{2}+(h/1000)^{2}}.
\end{align}
With a simplified path-loss-plus-noise model, the receive SNR and the bottleneck Shannon capacity are
\begin{align}
\mathrm{SNR}_i &= \frac{p_{tx}}{0.15+d_i^{2}}, \\
C(\mathbf{x}) &= B\cdot\log_2\!\left(1+\min(\mathrm{SNR}_1,\mathrm{SNR}_2)\right) \text{ Mbps}.
\end{align}
The service cost
\begin{equation}
S(\mathbf{x}) = 10\,p_{tx} + 0.06\,h + 0.8\,B
\end{equation}
captures the operating expense of higher transmit power, higher altitude (energy and ATC clearance) and wider bandwidth.

\paragraph{Constraints} The deployment must satisfy
\begin{align}
g_1 &: 25 - C(\mathbf{x}) \le 0, \\
g_2 &: (60+18\,d_h^{\max}) - h \le 0, \\
g_3 &: p_{tx}\cdot B - 90 \le 0, \\
g_4 &: d_{h,1} - 9 \le 0,\quad g_5: d_{h,2} - 9 \le 0,
\end{align}
with $d_h^{\max}=\max(d_{h,1},d_{h,2})$ corresponding to minimum capacity, line-of-sight clearance, power-bandwidth budget, and horizontal range, respectively. Following \cite{wei2026nawoa,li2026asksssa}, we adopt a quadratic penalty function with factor $\rho=10^{4}$:
\begin{equation}
y(\mathbf{x}) = -C(\mathbf{x}) + \frac{S(\mathbf{x})}{10} + \rho\sum_{k=1}^{5}\max(g_k,0)^{2}.
\label{eq:y}
\end{equation}
The optimizer minimizes \eqref{eq:y} so that lower scores correspond to higher capacity, lower service cost and smaller constraint violations.

\section{The Proposed PWOA}
\label{sec:pwoa}

\subsection{Standard WOA recap}
The WOA \cite{mirjalili2016woa} models bubble-net hunting by humpback whales through three operators: encircling, spiral updating and random search. Let $\mathbf{X}_i^{(t)}$ be the position of whale $i$ at iteration $t$ and $\mathbf{X}^{*,(t)}$ the best agent so far. With $a$ linearly decreasing from $2$ to $0$, $A=2ar_1-a$, $C=2r_2$, $r_1,r_2\sim U(0,1)$, $p\sim U(0,1)$, the update rule is
\begin{equation}
\mathbf{X}_i^{(t+1)}\!=\!
\begin{cases}
\mathbf{X}^{*}\!-\!A\!\cdot\!|C\mathbf{X}^{*}\!-\!\mathbf{X}_i|, & p\!<\!0.5,|A|\!<\!1 \\
\mathbf{X}_r\!-\!A\!\cdot\!|C\mathbf{X}_r\!-\!\mathbf{X}_i|, & p\!<\!0.5,|A|\!\ge\!1 \\
|\mathbf{X}^{*}\!-\!\mathbf{X}_i|\,e^{l}\!\cos(2\pi l)\!+\!\mathbf{X}^{*}, & p\!\ge\!0.5
\end{cases}
\label{eq:woa}
\end{equation}
where $l\sim U(-1,1)$. The linear decay of $a$ together with random initialization is the main source of premature convergence on highly multi-modal, constrained problems such as the UAV deployment in Section~\ref{sec:problem}.

\subsection{Improvement 1: Good Nodes Set initialization}
\begin{figure}[t]
\centering
\includegraphics[width=0.9\columnwidth]{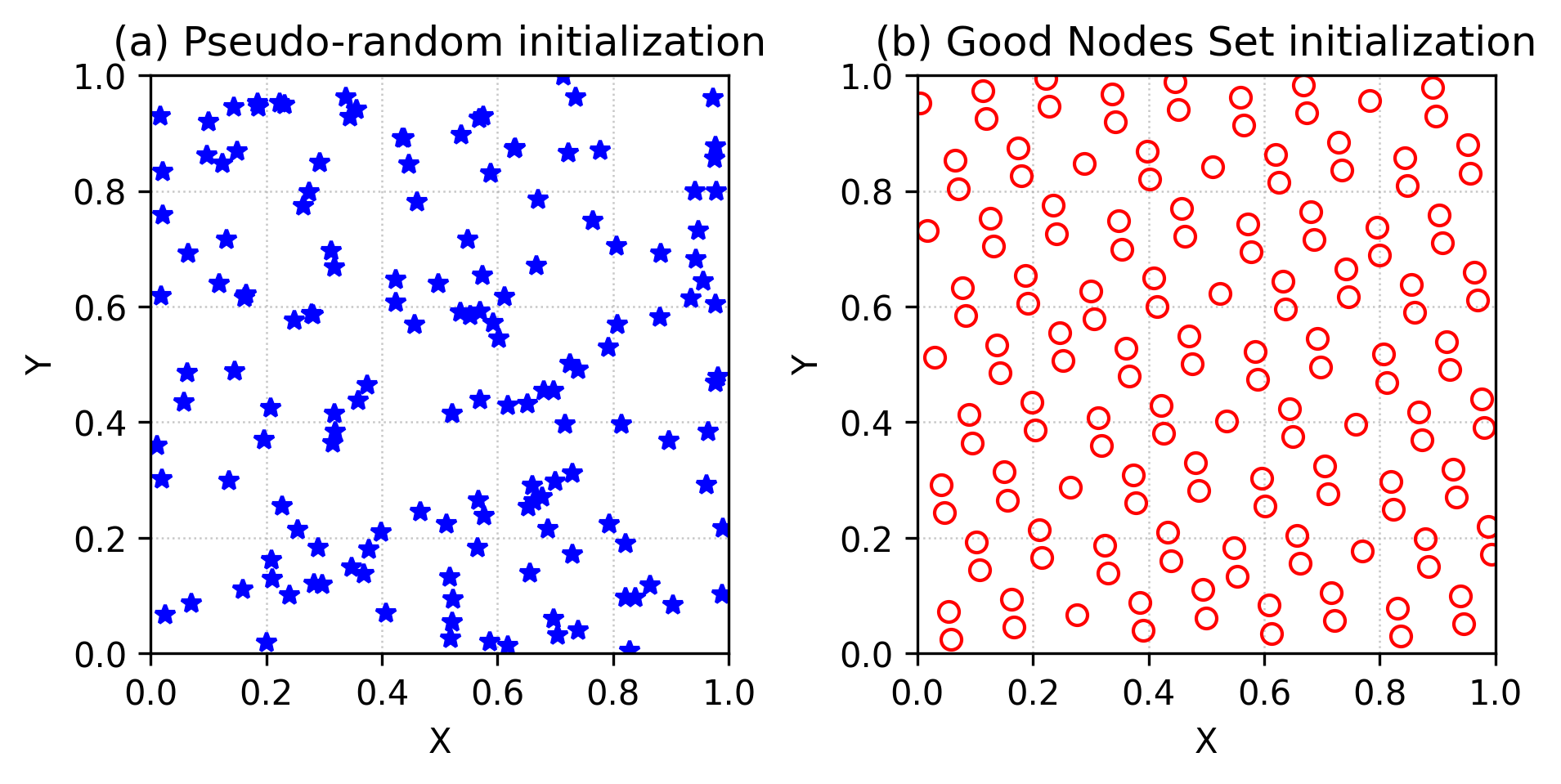}
\caption{2D comparison of pseudo-random initialization (left, blue stars) and Good Nodes Set initialization (right, red circles) for $N=150$ on $[0,1]^{2}$.}
\label{fig:init}
\end{figure}

Random initialization frequently leaves large regions of the search space empty, especially in the low-dimensional constrained setting where the feasible region is narrow. We adopt a deterministic Good Nodes Set initialization with a small jitter that preserves diversity across runs:
\begin{equation}
\mathbf{X}_{ij}^{(0)} = lb_j + \frac{ub_j - lb_j}{N-1}\cdot(i-1) + \varepsilon\,(ub_j-lb_j)\,(\xi_{ij}-0.5)
\label{eq:gns}
\end{equation}
for $i=1,\ldots,N$, $j=1,\ldots,D$, with $\xi_{ij}\sim U(0,1)$ and $\varepsilon=0.01$. This guarantees uniform coverage along every dimension and reduces the variance of the early-stage best fitness across independent runs. Fig.~\ref{fig:init} contrasts $150$ samples drawn by pseudo-random and Good Nodes Set initialization on the unit square; the latter exhibits visibly better space-filling and avoids the clumping and empty regions characteristic of random sampling.

\subsection{Improvement 2: Polynomial nonlinear decay of $a$}
The original linear schedule pushes the algorithm into exploitation as soon as half of the iterations are spent. Inspired by the parameter scheduling analysis of \cite{wei2025tswoa,wei2026geometric}, PWOA replaces the linear schedule with a polynomial nonlinear decay
\begin{equation}
a(t) = a_{\max} - (a_{\max}-a_{\min})\cdot\left(\frac{t}{T-1}\right)^{\mu},
\label{eq:poly}
\end{equation}
where $a_{\max}=2$, $a_{\min}=0$, and $\mu=2$. With $\mu>1$ the factor $a$ decreases slowly during the first half of the run, retaining the random-search branch ($|A|\ge 1$) longer and improving exploration; in the second half $a$ falls quickly so that the spiral and encircling branches dominate, yielding fast exploitation.

\subsection{Improvement 3: Stagnation-triggered POBL with elite local search}
\begin{figure}[t]
\centering
\includegraphics[width=0.9\columnwidth]{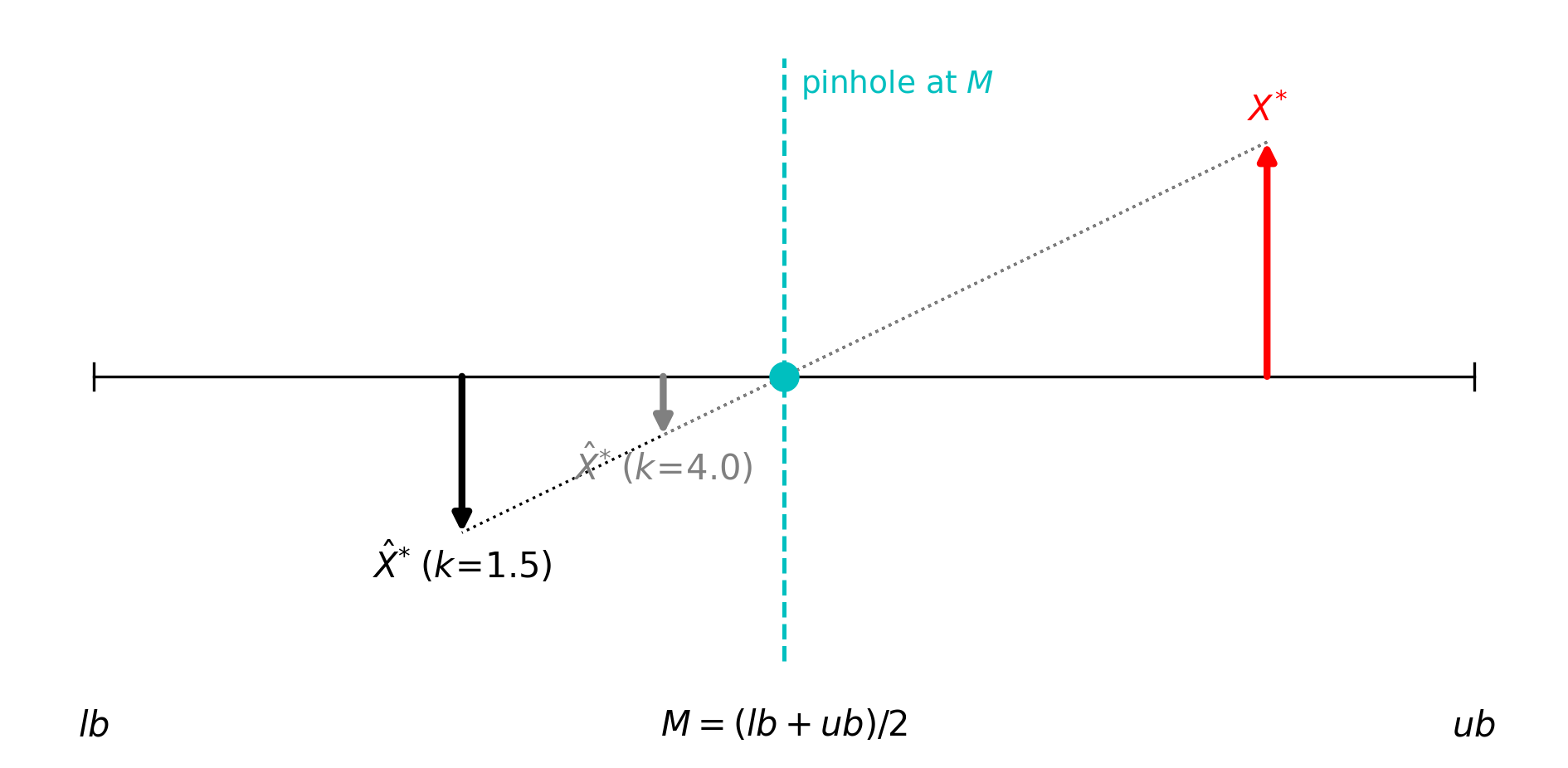}
\caption{Pinhole-imaging POBL geometry. The leader $X^{*}$ is reflected through the search-space centre $M$ to produce a candidate $\hat{X}^{*}$ whose magnification is $1/k(t)$: a small $k$ (early iterations) yields a long jump (black) and a large $k$ (late iterations) yields a short jump (grey).}
\label{fig:pobl}
\end{figure}

The third innovation tackles local-optima trapping. PWOA monitors a stagnation counter $\sigma$ of the global-best score. (a) At every iteration we first perform an \emph{elite Gaussian local search} around the leader,
\begin{equation}
\tilde{\mathbf{X}}^{*} = \mathbf{X}^{*} + \boldsymbol{\eta}\odot(\mathbf{ub}-\mathbf{lb}),\quad \boldsymbol{\eta}\sim\mathcal{N}(0,\beta(t)\mathbf{I}),
\label{eq:elite}
\end{equation}
with shrinking step $\beta(t)=\beta_0(1-t/(T-1))$ and $\beta_0=0.05$. The candidate replaces the leader only when it is strictly better. (b) Once $\sigma$ exceeds the threshold $\sigma_{\max}=15$, we apply a pinhole-imaging opposition-based learning (POBL) transform with progress-adaptive imaging factor $k(t)$:
\begin{equation}
\hat{\mathbf{X}}^{*}_{j} = \frac{lb_j+ub_j}{2}+\frac{ub_j-lb_j}{2k(t)}-\frac{X^{*}_{j}}{k(t)},
\label{eq:pobl}
\end{equation}
where $k(t)=1.5+(4.0-1.5)\cdot t/(T-1)$. A small $k$ at the beginning generates an aggressive long-range jump to escape early local minima; a large $k$ at the end maps the candidate close to the centroid of the search space, enabling controlled diversification near termination. The candidate is always injected into the worst individual, breaking population stagnation, and replaces the leader if it is strictly better. The counter $\sigma$ is then reset. Fig.~\ref{fig:pobl} illustrates the pinhole-imaging geometry: the leader $X^{*}$ is the object, $M=(lb+ub)/2$ acts as the pinhole, and the candidate $\hat{X}^{*}$ is the inverted image whose distance from $M$ shrinks as $k$ grows.

\subsection{PWOA pseudocode and complexity}
The complete procedure is summarized in Algorithm~\ref{alg:pwoa}. Per iteration PWOA performs $\mathcal{O}(N\cdot D)$ position updates plus a constant number of leader evaluations, so the overall time complexity is $\mathcal{O}(N\cdot D\cdot T)$, identical to standard WOA.

\begin{algorithm}[t]
\caption{PWOA}
\label{alg:pwoa}
\begin{algorithmic}[1]
\State Initialize $\mathbf{X}^{(0)}$ via GNS \eqref{eq:gns}; evaluate fitness; $\mathbf{X}^{*}\!\gets\!\arg\min$; $\sigma\!\gets\!0$.
\For{$t=0,1,\ldots,T-1$}
    \State Compute $a(t)$ via \eqref{eq:poly}.
    \For{$i=1,\ldots,N$}
        \State Update $\mathbf{X}_i^{(t+1)}$ via \eqref{eq:woa}; clip and evaluate.
        \State If improved, update $\mathbf{X}^{*}$.
    \EndFor
    \State Generate $\tilde{\mathbf{X}}^{*}$ via \eqref{eq:elite}; if better, accept.
    \State Update $\sigma$.
    \If{$\sigma\ge\sigma_{\max}$}
        \State Generate $\hat{\mathbf{X}}^{*}$ via \eqref{eq:pobl}; if better, accept.
        \State Inject $\hat{\mathbf{X}}^{*}$ into worst individual; $\sigma\gets 0$.
    \EndIf
\EndFor
\State \Return $\mathbf{X}^{*}$, fitness curve.
\end{algorithmic}
\end{algorithm}

\section{Experiments}
\label{sec:exp}

\subsection{Setup}
We compare PWOA with the original WOA \cite{mirjalili2016woa}, SCA \cite{mirjalili2016sca} and an improved PSO (IPSO) variant \cite{wei2024apupso} that uses a Tent-map initialization, Levy-flight perturbation and a self-adaptive Student-$t$ distribution operator. All algorithms are run on the deployment problem of Section~\ref{sec:problem} with population size $N=30$, maximum iterations $T=500$ and $30$ independent runs (seeds 0--29). Each run uses a different random seed but the seed sequence is shared across algorithms to ensure fairness. PWOA-specific hyper-parameters are set to $\mu=2$, $\sigma_{\max}=15$, $k_{\min}=1.5$, $k_{\max}=4$, $\beta_0=0.05$. Experiments were performed on a Linux workstation in Python~3.13 / NumPy~2.x.

\subsection{Statistical results}
Table~\ref{tab:res} reports the Best, Worst, Mean and standard deviation (Std) of the final fitness over $30$ runs. Lower is better. PWOA attains the lowest Best, lowest Worst, lowest Mean and lowest Std simultaneously, indicating that it is both more accurate and more reliable than all three baselines. In particular, PWOA reduces the Mean by $2.9\%$ over WOA, $1.4\%$ over SCA and $18.5\%$ over IPSO, while reducing the Std by $63\%$ over WOA and $87\%$ over IPSO.

\begin{table}[t]
\centering
\caption{Results over 30 independent runs ($N=30$, $T=500$). Lower is better for all metrics; the best value in each column is bolded.}
\label{tab:res}
\setlength{\tabcolsep}{4pt}
\resizebox{\columnwidth}{!}{%
\begin{tabular}{lcccc}
\toprule
Algorithm & Best $\downarrow$ & Worst $\downarrow$ & Mean $\downarrow$ & Std $\downarrow$ \\
\midrule
\textbf{PWOA} & $\bm{3.317\times 10^{6}}$ & $\bm{3.512\times 10^{6}}$ & $\bm{3.412\times 10^{6}}$ & $\bm{5.37\times 10^{4}}$ \\
WOA  & $3.335\times 10^{6}$ & $3.934\times 10^{6}$ & $3.515\times 10^{6}$ & $1.44\times 10^{5}$ \\
SCA  & $3.324\times 10^{6}$ & $3.605\times 10^{6}$ & $3.459\times 10^{6}$ & $6.30\times 10^{4}$ \\
IPSO & $3.477\times 10^{6}$ & $4.821\times 10^{6}$ & $4.185\times 10^{6}$ & $4.09\times 10^{5}$ \\
\bottomrule
\end{tabular}
}
\end{table}

\subsection{Convergence behaviour}
\begin{figure}[t]
\centering
\includegraphics[width=0.9\columnwidth]{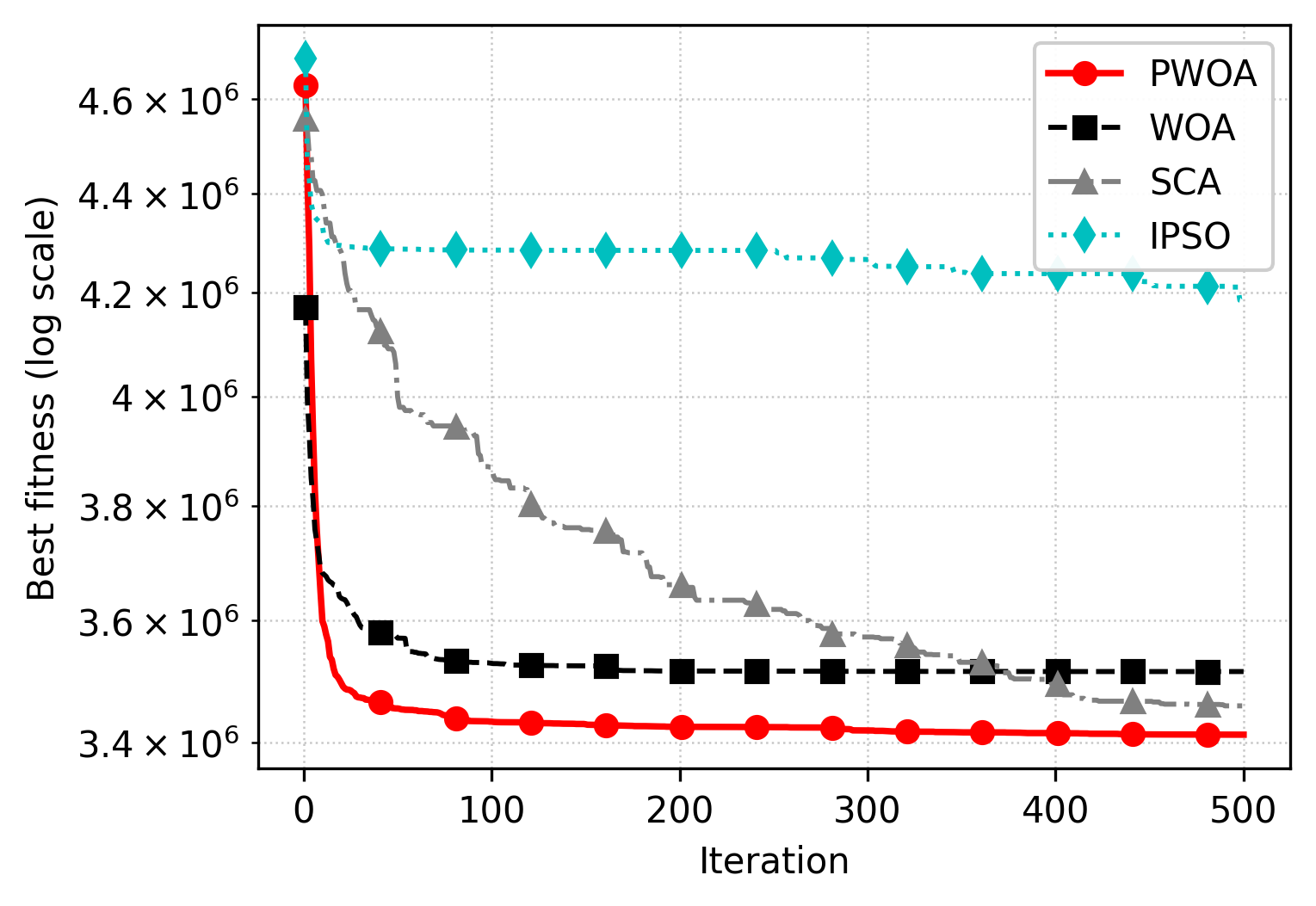}
\caption{Average convergence curves of the four algorithms on the UAV relay deployment problem (30 runs, log-scale $y$-axis).}
\label{fig:curve}
\end{figure}

Fig.~\ref{fig:curve} shows the average best-so-far fitness over the 30 runs (logarithmic $y$-axis). PWOA falls below the WOA curve within the first 30 iterations and stays the lowest for the remaining 470 iterations, confirming that the GNS initialization and the polynomial schedule jointly accelerate the early descent. SCA initially stagnates and only catches up close to iteration 500, while IPSO is the slowest because the Student-$t$ operator pushes the swarm too far from the feasible region in early iterations.

\subsection{Discussion}
The three improvements form a coverage--transition--recovery chain. GNS reduces sampling discrepancy and supplies a more balanced population for WOA's encircling and spiral operators. The polynomial schedule retains larger $|A|$ values early and accelerates exploitation only after sufficient exploration. Elite local search and POBL are conditional recovery mechanisms tied to leader stagnation rather than always-on perturbations. Thus, the modules are necessary for different failure modes: missing feasible regions, exploiting too early, and remaining trapped after population contraction. This coupling explains the tighter final distribution in Table~\ref{tab:res}, where PWOA reaches a Std of $5.37\times10^{4}$.

\section{Conclusion}
\label{sec:conclusion}
We have presented PWOA, a Polynomial-decay and Pinhole-imaging Whale Optimization Algorithm tailored to the UAV relay communication deployment problem. PWOA augments standard WOA with a Good Nodes Set initialization, a polynomial nonlinear convergence-factor schedule and a stagnation-triggered POBL operator combined with elite Gaussian local search. Across 30 independent runs of a five-dimensional, five-constraint deployment task, PWOA outperforms WOA, SCA and an improved PSO baseline on all four reported metrics (Best, Worst, Mean and Std) and exhibits the fastest average convergence. Future work will extend PWOA to multi-UAV cooperative deployment with mobile users and energy budgets, and to closed-loop scenarios where channel statistics are estimated online.

\section{Acknowledgment}
This work was supported by the Macao Science and Technology Development Fund (FDCT-MOST: 0018/2025/AMJ) and Macao Polytechnic University (MPU grant: RP/FCA-01/2025).

\end{document}